\documentclass[%
 reprint,
superscriptaddress,
 amsmath,amssymb,
prl,
floatfix,
]{revtex4-1}

\usepackage{braket}				% Allows Dirac notation
\usepackage{graphicx}			% Include figure files
\usepackage{bm}				% bold math
\usepackage[usenames, dvipsnames]{color}
\usepackage{xcolor}	
\usepackage{hyperref}			% add hypertext capabilities
\hypersetup{
    colorlinks,
    linkcolor={blue!50!black},
    citecolor={blue!50!black},
    urlcolor={blue!80!black}
}
\usepackage[caption=false]{subfig}	
\usepackage{mathtools}
\usepackage{soul} 				% for crossing things out

\newcommand{\abs}[1]{\left| #1 \right|} 	% for absolute value
\DeclareMathOperator{\tr}{tr} 			% for trace

\DeclareMathOperator\arccosh{arccosh} 	% Inverse hyperbolic cosine
\newcommand{\nn}{\nonumber}		% for no numbering

\usepackage{colonequals} %Get a nice looking :=
\newcommand{\ce}{\colonequals}

\newcommand{\tcb}{\textcolor{blue}}			
\newcommand{\tcr}{\textcolor{red}}
\newcommand{\tcp}{\textcolor{purple}}			
\newcommand{\tco}{\textcolor{orange}}

\begin{document}

%\preprint{APS/123-QED}

\title{Harvesting Entanglement from the Black Hole Vacuum}

\author{L. J. Henderson}
\email[]{l7henderson@uwaterloo.ca}
\affiliation{Institute for Quantum Computing, University of Waterloo, Waterloo, Ontario, Canada, N2L 3G1}
\affiliation{Department of Physics and Astronomy, University of Waterloo, Waterloo, Ontario, Canada, N2L 3G1}

\author{R. A. Hennigar}
\email[]{rhennigar@uwaterloo.ca}
\affiliation{Department of Physics and Astronomy, University of Waterloo, Waterloo, Ontario, Canada, N2L 3G1}

\author{R. B. Mann}
\email[]{rbmann@uwaterloo.ca}
\affiliation{Department of Physics and Astronomy, University of Waterloo, Waterloo, Ontario, Canada, N2L 3G1}
\affiliation{Institute for Quantum Computing, University of Waterloo, Waterloo, Ontario, Canada, N2L 3G1}
\affiliation{Perimeter Institute for Theoretical Physics, 31 Caroline St. N., Waterloo, Ontario, Canada, N2L 2Y5}

\author{A. R. H. Smith}
\email[]{alexander.r.smith@dartmouth.edu}
\affiliation{Department of Physics and Astronomy, Dartmouth College, Hanover, New Hampshire 03755, USA}

\author{J. Zhang}
\email[]{jialinzhang@hunnu.edu.cn \\ }
\affiliation{Department of Physics and Synergetic Innovation Center for Quantum Effects and Applications, Hunan Normal University, Changsha, Hunan 410081, China}

\date{\today}	% It is always \today, today,

\begin{abstract}
We implement the entanglement harvesting protocol, in which two Unruh-DeWitt detectors become entangled through local interactions with a quantum field, for the first time in the vicinity of a black hole. Our study, focusing on the BTZ black hole, reveals that black holes inhibit entanglement harvesting. The entanglement harvested rapidly falls to zero when two  detectors with fixed proper separation approach the horizon. This effect is a combination of black hole radiation and gravitational red shift, both generic properties of horizons, {\mbox{suggesting it is a general result for black holes.}}
\end{abstract}

\maketitle

%========================================
%========================================

In recent years, there has been considerable interest in the role played by entanglement in quantum field theories.
 This research has drawn motivation from disparate areas of physics including  critical phenomena in condensed matter systems \cite{Osterloh:2002, Vidal:2003, Amico:2008}, in describing  non-classical states of light  \cite{Mandel:1995, Quantum-Information:2011}, in explaining the origin of black hole entropy \cite{Bombelli:1986, Callan:1994, Srednicki:1993}, and perhaps most spectacularly, in the anti-de Sitter/conformal field theory correspondence, where the entanglement entropy associated with a region of a conformal field theory is related to minimal surfaces in a bulk gravitational theory~\cite{Ryu:2006}.

Within the framework of algebraic quantum field theory, Summers and Werner \cite{Summers:1985, *Summers:1987fn, *Summers:1987} demonstrated that the vacuum state of a free quantum field in Minkowski space, as seen by local inertial observers, is entangled, and that the correlations seen by these observers are strong enough to violate Bell-type inequalities, even if the observers are in spacelike separated regions.  This result is surprising\,---\,it suggests that no source of entanglement is necessary to detect a violation of Bell's inequality, the observation of vacuum fluctuations suffices.

An operational approach to the study of the entanglement structure of a quantum field theory was introduced in 1991 by Valentini \cite{Valentini:1991}. He showed that two initially uncorrelated atoms, which interact locally with the electromagnetic vacuum, can exhibit nonlocal correlations even if they remain spacelike separated throughout the interaction. This implies that any entanglement that results between the atoms must have been transferred from entanglement already present in the electromagnetic vacuum. Quantifying the resulting entanglement between the atoms thus provides an indicator of  electromagnetic vacuum entanglement in the regions where the atoms were located. In 2002 Reznik \emph{et al.} \cite{Reznik:2002fz,Reznik:2005} demonstrated a similar effect using two Unruh-DeWitt detectors interacting locally with the vacuum state of a scalar field.

This process of localized detectors extracting entanglement/nonlocal correlations from the vacuum state of a quantum field has since become known as entanglement harvesting \cite{Salton:2014jaa}, and has been studied in a variety of different situations, ranging from the extraction of resources from the vacuum \cite{Martin-MartinezSUS}, to entanglement generation between hydrogen-like atoms \cite{Pozas-Kerstjens:2015, Pozas-Kerstjens:2016}, and has been shown to depend on the underlying spacetime geometry \cite{Steeg:2009} and topology \cite{Smith:2016a}. The entanglement harvesting protocol provides a simple operational means for extracting vacuum entanglement that is immediately applicable to quantum field theories on curved spacetimes.

Here we report on the first investigation of entanglement harvesting in a black hole background. We present a general formalism, valid for any stationary spacetime, then specialize to a conformally coupled scalar field on the BTZ black hole. The key finding of our study is that \textit{black holes inhibit entanglement harvesting}. Due to a combination of red shift and the Hawking effect, the entanglement harvested falls rapidly to zero when two detectors with fixed proper separation approach the horizon. This cannot be evaded, even if the detectors are timelike separated. Since Hawking radiation and red shift are generic properties of black hole horizons, we expect this result to be general.

As a simplified model of an atom interacting with the vacuum, we employ the Unruh-DeWitt detector \cite{Unruh:1976, DeWitt:1979}, which consists of a two-level quantum system moving along the spacetime trajectory $x_D(\tau)$, parametrized by the detector's proper time $\tau$ and interacting locally with a scalar field $\phi(x)$.  The ground and excited states of the detector are denoted as $\ket{0}_D$ and $\ket{1}_D$, respectively, and separated by an energy gap $\Omega_D$. In the interaction picture, the Hamiltonian describing the interaction of the detector with the field is
\begin{align}
H_D(\tau) &= \lambda \chi_D\! \left(\tau \right)\Big(e^{ i\Omega \tau} \sigma^+  +  e^{- i\Omega \tau}\sigma^- \Big) \otimes  \phi\left[x_D(\tau)\right] \label{InteractionHamiltonian}
\end{align}
where $\chi_D(\tau) \leq 1 $ is a switching function controlling the duration of the interaction, and $\sigma^+ \ce  \ket{1}_D\!\bra{0}_D$ and $\sigma^- \ce  \ket{0}_D\!\bra{1}_D$ are ladder operators acting on the Hilbert space associated with the detector. Although simple, this model captures the relevant features of the light-matter interaction when no angular momentum exchange is involved \cite{Martin-Martinez2013, Alvaro}.

Consider two detectors, $A$ and $B$, with trajectories $x_A(\tau_A)$ and $x_B(\tau_B)$ parametrized by the detectors' proper times $\tau_A$ and $\tau_B$, respectively. Suppose these detectors are initially ($\tau_A, \tau_B \to -\infty$) prepared in their ground state, and the state of the field is in an appropriately defined vacuum state $\ket{0}$, so that the joint state of the detectors and field together is $\ket{\Psi_i} = \ket{0}_A \ket{0}_B \ket{0}$. Given that the interaction between each detector and the field is described by the Hamiltonian in \eqref{InteractionHamiltonian}, which for detectors $A$ and $B$ we denote as $H_A$ and $H_B$, respectively,  the final ($\tau_A, \tau_B \to \infty$) state of the detectors is given by
\begin{align}
\ket{\Psi_f} = \mathcal{T} e^{  -i \int  dt\, \left[ \frac{d \tau_A}{dt} H_A(\tau_A) + \frac{d \tau_B}{dt} H_B(\tau_B) \right] } \ket{\Psi_i}, \label{totalfinalstate}
\end{align}
where $\mathcal{T}$ is the time ordering operator and we have chosen to evolve the field and detectors with respect to an appropriate coordinate time $t$ with respect to which the vacuum state of the field is defined.

The final state of the detectors alone is obtained from~\eqref{totalfinalstate} by tracing out the field degrees of freedom, $\rho_{AB} \ce \tr_\phi \big( \ket{\Psi_f}\!\bra{\Psi_f} \big)$, which to leading order in the interaction strength, in the basis $\{ \ket{0}_A \ket{0}_B, \ket{0}_A \ket{1}_B, \ket{1}_A \ket{0}_B, \ket{1}_A \ket{1}_B \}$, is given by~\cite{Smith:2016a, Smith:2017b}
\begin{align}
\rho_{AB} &= \begin{pmatrix}
1 - P_A - P_B  & 0 & 0 & X \\
0 & P_B  & C & 0 \\
0 & C^* & P_A & 0 \\
X^* & 0 & 0 & 0
\end{pmatrix} + \mathcal{O}\!\left(\lambda^4\right), \label{FinsalState2}
\end{align}
where
\begin{widetext}
\begin{align}
&P_D \ce \lambda^2 \int d\tau_D  d \tau_D' \, \chi_D(\tau_D) \chi_D(\tau_D') e^{-i \Omega_D \left(\tau_D-\tau_D'\right)} W\!\left(x_D(t) ,x_D(t')\right) \label{PJ}  \quad \mbox{for} \quad D \in \{A,B\},
 \\
&C\ce \lambda^2 \int d\tau_A d\tau_B   \, \chi_A(\tau_A) \chi_B(\tau_B) e^{- i \left( \Omega_A \tau_A - \Omega_B \tau_B \right)} W\!\left(x_A(t) , x_B(t')\right) \label{defC}, \\
&X \ce-\lambda^2  \int  d\tau_A d\tau_B \,   \chi_A(\tau_A) \chi_B(\tau_B)  e^{-i\left( \Omega_A  \tau_A + \Omega_B  \tau_B\right)}
\Big[ \theta(t'-t) W\!\left(x_A(t), x_B(t')\right)  + \theta(t-t') W\!\left(x_B(t'),x_A(t) \right)  \Big] , \label{defX}
\end{align}
\end{widetext}
where $W(x,x')\ce\bra{0} \phi(x) \phi(x') \ket{0}$ is the Wightman function associated with the field and $\theta(\,\cdot\,)$ is the Heaviside function. Note that in \eqref{defC} and \eqref{defX} $\tau_A = \tau_A(t)$ and $\tau_B = \tau_B(t')$ are functions of the coordinate $t$ (and so  $d\tau_D =\frac{d\tau_D}{dt} dt$).
Analysis of the next to leading order contribution to $\rho_{AB}$ demonstrates that the leading order contribution to $\rho_{AB}$ is completely positive \cite{Smith:2017b}.

The reduced states of the individual detectors are
\begin{align}
\rho_D &\ce\begin{pmatrix}
1-P_D & 0 \\
0 & P_D
\end{pmatrix}
+\mathcal{O}\!\left( \lambda^4\right) \quad \mbox{for} \quad D \in \{A,B\} ,
\label{reducedD}
\end{align}
which justifies interpreting $P_A$ and $P_B$ as the transition probability that either detector $A$ or $B$ will become excited as a result of its interaction with the field. As the matrix elements $C$ and $X$ do not appear in the reduced states of either detector, they are associated with correlations and entanglement shared between the detectors.

To quantify the entanglement harvested by the detectors, we use the concurrence as an entanglement measure~\footnote{Other literature on entanglement harvesting has used the negativity as a measure of entanglement, which when evaluated for the state given in \eqref{FinsalState2} to leading order yields
{\scriptsize
\begin{align*}
\mathcal{N}(\rho_{AB}) &=  \max \!\left[ \, 0, \,   \sqrt{ \abs{X}^2 + \left(\frac{P_A-P_B}{2}\right)^2} -\frac{P_A+P_B}{2}\, \right] .
\end{align*} }
However, unlike  concurrence, the negativity is not a simple difference between a nonlocal and a local term. It is for this reason we employ the concurrence in this article.}, which for \eqref{FinsalState2} is~\cite{Wootters:1997id, Smith:2016a, Smith:2017b}
\begin{align}
\mathcal{C}(\rho_{AB})= 2 \max \!\left[ \, 0, \ \abs{X} -  \sqrt{P_A P_B}\, \right]  + \mathcal{O}\!\left(\lambda^4\right)
\label{concurrence}.
\end{align}
Being a simple difference of a local and non-local terms, $\mathcal{C}(\rho_{AB})$ is convenient in interpreting the results to follow.

The amount of entanglement harvested depends on the detectors' energy gaps, switching functions, and trajectories. However, if one fixes these quantities and compares the entanglement harvested for different spacetime parameters (e.g. horizon radius) or the location of the   detectors in a given spacetime, then the difference in the amount of entanglement harvested can be attributed to  the properties of the spacetime.

In (2+1)-dimensions, the Einstein equations with a negative cosmological constant $\Lambda = -1/\ell^2$ admit the BTZ black hole solution \cite{Banados:1992,Banados:1993}
\begin{align}
&ds^2 = - \left( \frac{r^2-r_h^2}{\ell^2}\right) dt^2 + \left(  \frac{\ell^2}{r^2-r_h^2}\right)  dr^2 + r^2 d \phi^2 ,\label{BTZmetric}
\end{align}
expressed here in Schwarzchild-like coordinates, $t \in (-\infty, \infty)$, $r\in(0,\infty)$, and  $\phi \in (0, 2 \pi)$. This solution has a horizon at $r_h \ce \ell \sqrt{M}$, where $M$ is the black hole mass, and is asymptotic to anti de Sitter space (AdS$_3$).
\begin{figure}[t]
\includegraphics[width = 0.4\textwidth]{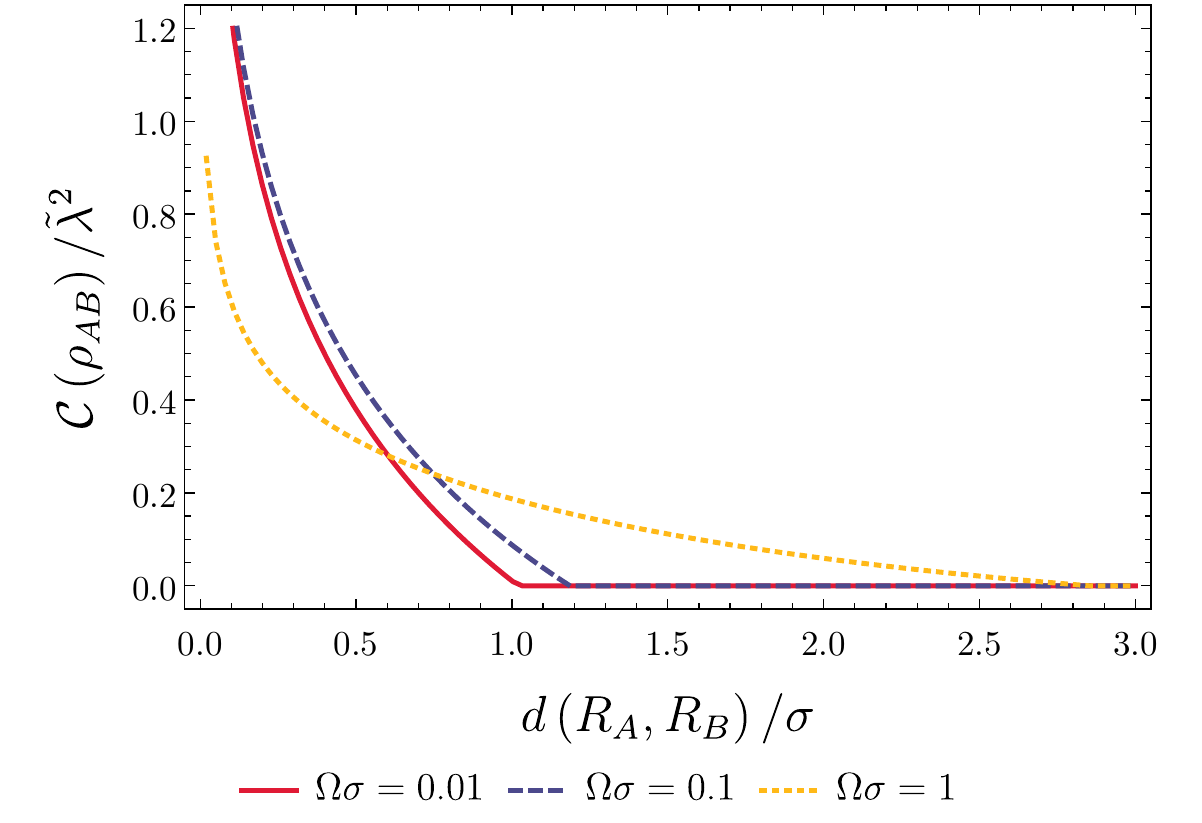}
\vspace{0.05in}
\caption{%
The concurrence ${\cal C}(\rho_{AB})/ \tilde{\lambda^2}$ between two detectors in the exterior region of the BTZ black hole is plotted as a function of the proper distance separating them; different energy gaps $\Omega\sigma $ of the detectors are shown and $\tilde{\lambda}\ce\lambda\sqrt{\sigma}$ denotes the dimensionless coupling strength. The proper distance between detector $A$ and the horizon is $d(r_h, R_A)/\sigma =1$, and $\ell/\sigma = 10$ and $M=1$.
}
\label{fig:BTZ1}
\end{figure}

The advantage of considering detectors in the BTZ spacetime \cite{Hodfkinson:2012,Hodfkinson:2012a,Smith:2014}
is that  for  a conformally coupled scalar field (in the Hartle-Hawking vacuum) the Wightman functions are known analytically \cite{Lifschytz:1994,Carlip:2003}.
Since the BTZ spacetime may be constructed by a topological identification $\Gamma$ of AdS$_3$, the Wightman function associated with the Hartle-Hawking vacuum  in the BTZ spacetime may be expressed as the image sum
\begin{align}
W_{\rm BTZ}(x,x') &= \sum_{n=-\infty}^\infty  \, W_{\rm AdS_3}(x, \Gamma^n x'), \label{BTZWightman}
\end{align}
where $W_{\rm AdS_3}(x, x')$ is the vacuum Wightman function associated with a massless conformally coupled scalar field in  AdS$_3$ and $\Gamma x'$ denotes the action of the identification  on the spacetime point $x'$. Explicitly, for two spacetime points $x$ and $x'$ outside the black hole horizon,
\begin{align}
W_{\rm {BTZ}}(x,    x') =  \frac{1}{4 \pi \sqrt{2} \ell } \sum_{n=-\infty}^\infty\left[ \frac{1}{\sqrt{\sigma_n}} -  \frac{\zeta}{\sqrt{\sigma_n + 2}} \right]  ,
\end{align}
where   $\zeta \in \{ -1, 0, 1 \}$ specifies either Neumann ($\zeta = -1$), transparent  ($\zeta = 0$), or Dirichlet  ($\zeta = 1$) boundary conditions satisfied by the field at spatial infinity (below we take $\zeta = 1$), and
\begin{align}
\sigma_n &\ce  \frac{r r'}{r_h^2} \cosh\! \left[\frac{r_h}{\ell} ( \Delta \phi - 2 \pi n) \right] -1 \nn \\
&\quad - \frac{\sqrt{ (r^2 - r_h^2)(r'^2 - r_h^2)}}{r_h^2}  \cosh \!\left[\frac{r_h}{\ell^2} \Delta t \right],   \label{BTZsigma1}
\end{align}
where $\Delta \phi \ce \phi-\phi'$ and $\Delta t \ce t-t'$.

Suppose detectors $A$ and $B$ are at fixed distances $R_A$ and $R_B$ outside the horizon of the BTZ black hole, $R_A, R_B >r_h$. The spacetime trajectories are
\begin{align}
&x_D(\tau_D) \ce \left\{ t =  \tau_D/\gamma_D,    r = R_D,  \phi=\Phi_D \right\}, \nonumber \\
&\gamma_D\ce\frac{\sqrt{R_D^2- r_h^2}}{\ell}, \quad \mbox{for} \quad D \in \{A,B\},
\end{align}
\label{2trajectories}
and without loss of generality we will set $R_A<R_B$.

We will choose the switching functions of the detectors to be Gaussian, $\chi_D(\tau_D) = \exp\left(-\tau_D^2/2\sigma^2\right)$,
with the interpretation that each detector interacts with the field for an approximate amount of proper time $k \sigma$, centred around the spacelike hypersurface $t=0$; {$k$ should be chosen so that at the proper time $k\sigma$ the detectors' interaction with the field is negligible}. In addition, we consider detectors with the same energy gap $\Omega_A = \Omega_B~=~\Omega$.

\begin{figure*}[t]
\includegraphics[width = 0.32\textwidth]{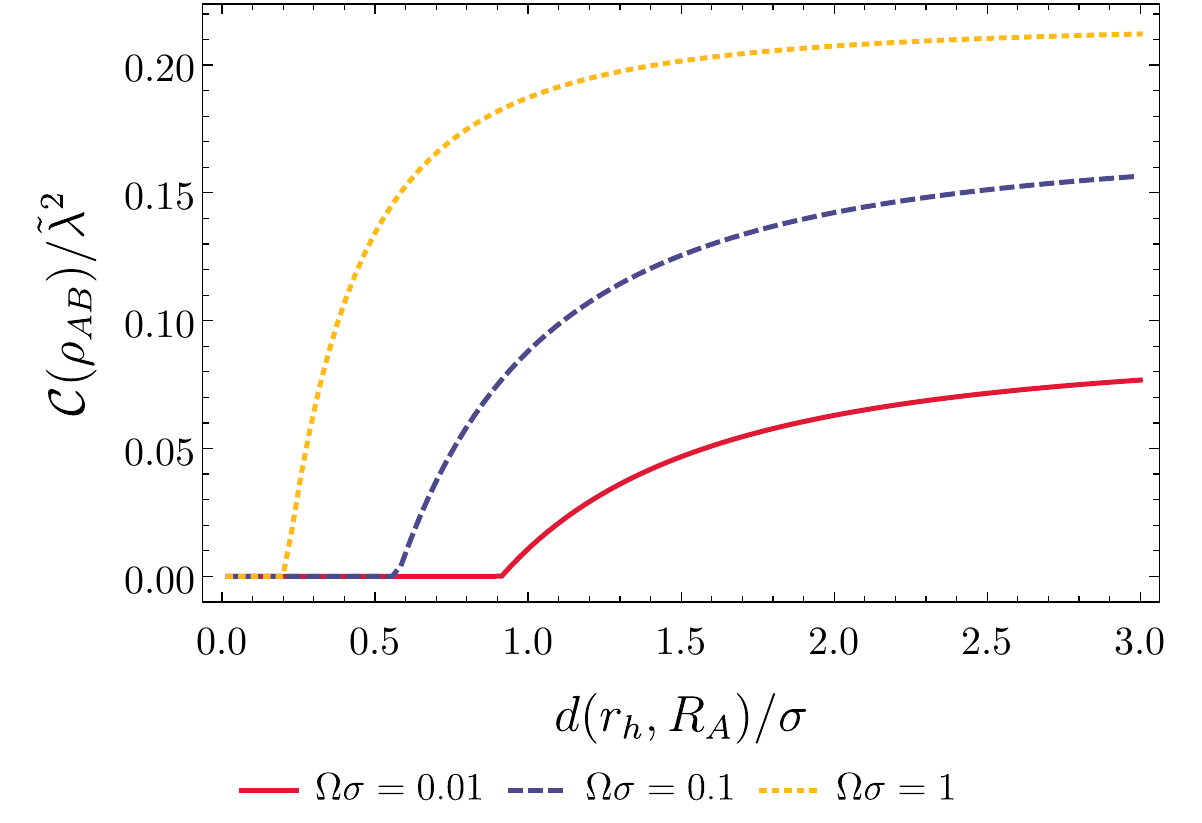}
\includegraphics[width = 0.32\textwidth]{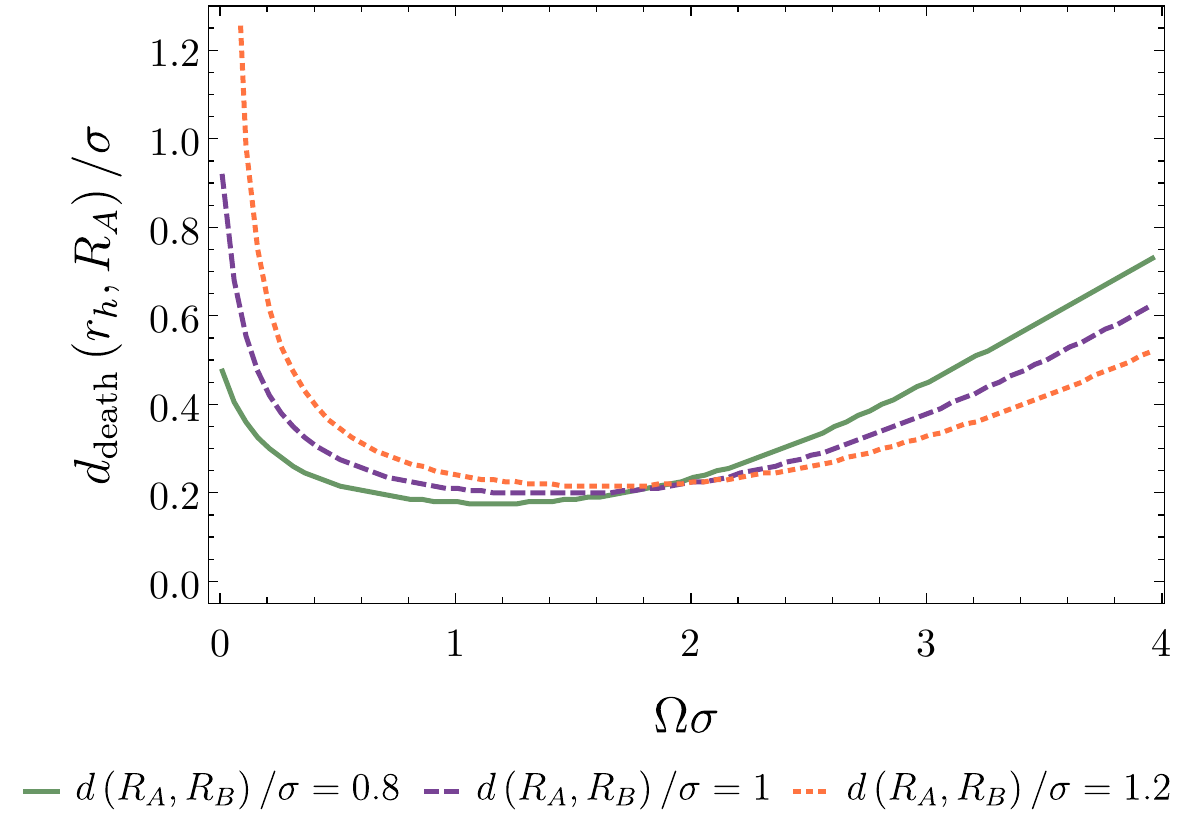}
\includegraphics[width = 0.32\textwidth]{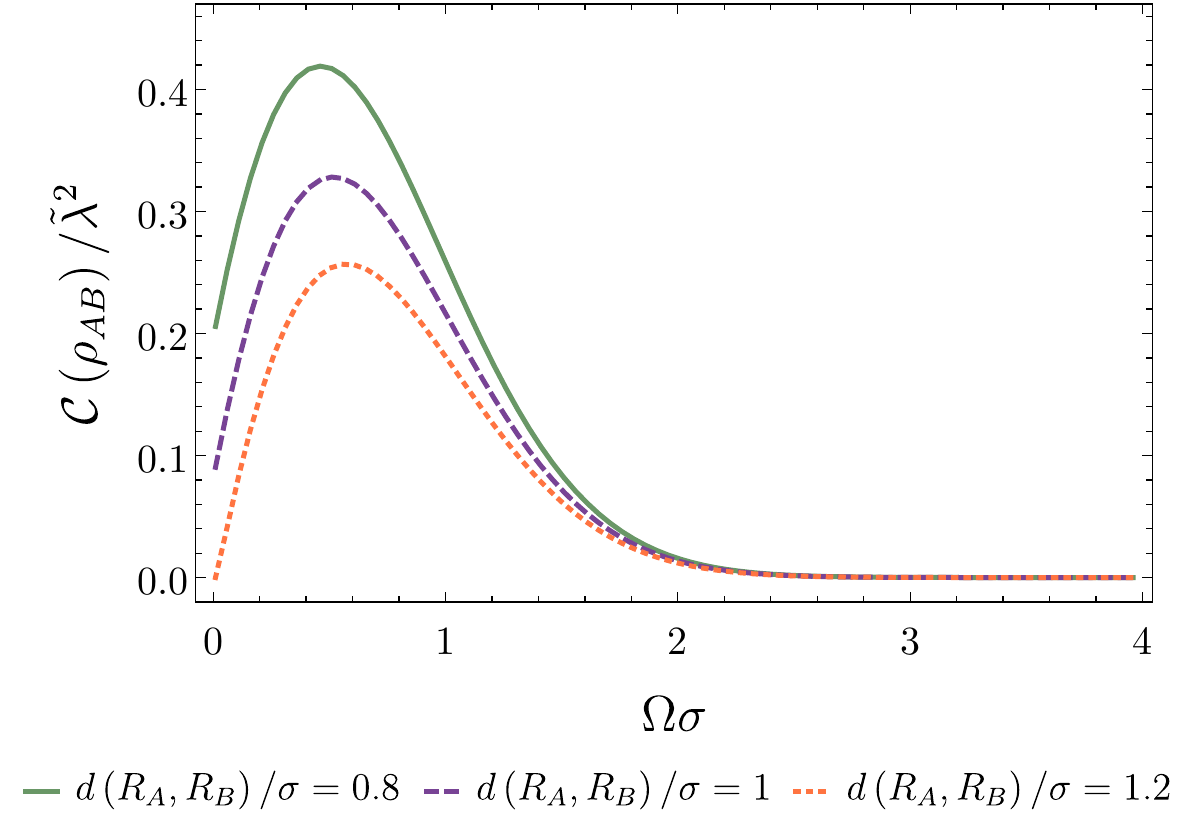}
\vspace{0.05in}
\caption{%
{\it Left}: The concurrence ${\cal C}(\rho_{AB})$ between two detectors in the exterior region of the BTZ black hole is plotted as a function of the proper distance detector $A$ is from the horizon; different energy gaps $\Omega \sigma $ of the detectors are shown. Here, the separation between the two detectors is fixed to be $d(R_A, R_B)/\sigma = 1$. {\it  Center}: A plot of the distance detector $A$ is from the horizon $d(r_h, R_A)$ when the entanglement vanishes as a function of the detectors' energy gap $\Omega \sigma$ is shown for three choices of detector separation. {\it  Right}: The  concurrence  in the large $d(r_h, R_A)$ limit is plotted as a function of the detectors' energy gap $\Omega\sigma$. In this plot, the concurrence has been evaluated at $d(r_h, R_A)/\sigma = 100$. In all plots, we have taken $\ell/\sigma = 10$ and $M=1$.
}
\label{fig:BTZ2}
\end{figure*}

Having specified the detector trajectories, switching functions, and energy gaps, we  compute numerically the concurrence ${\cal C}(\rho_{AB})$ of the final state of the two detectors, quantifying how much entanglement has been harvested from the field. (The explicit forms of the integrals are presented in the supplemental material.) We plot ${\cal C}(\rho_{AB})$ as a function of proper detector separation in Fig.~\ref{fig:BTZ1} and for fixed proper detector separation as a function of proper distance from the horizon in Fig.~\ref{fig:BTZ2}. Note the proper distance between two points $x_1=(t,R_1, \phi)$ and $x_2=(t,R_2, \phi)$ is
\begin{align}
d(R_1,R_2)
&= \ell \ln \left[ \frac{R_2+\sqrt{R_2^2-r_h^2}}{R_1+\sqrt{R_1^2-r_h^2}}\right],
\end{align}
where $R_2\geq R_1 \geq r_h$.

In Fig.~\ref{fig:BTZ1} we consider detector $A$ at a fixed proper distance from the black hole horizon and plot ${\cal C}(\rho_{AB})$ as a function of the proper distance between the two detectors. We observe that as the separation between the detectors grows, the entanglement between the detectors decreases. This behaviour is as expected since  correlations in the vacuum state are small for spacetime points separated by a large distance, which can be seen from the BTZ Wightman function in~\eqref{BTZWightman}. We also observe that the entanglement decreases more slowly for detectors with larger energy gap, but always vanishes for finite detector separation. This is a consequence of the fact that detectors with a larger gap are harder to excite by random fluctuations, so the correlations can dominate the transition probabilities of the detectors over larger separations; see~\eqref{concurrence}.

 \begin{figure*}[t]
\includegraphics[width = 0.3\textwidth]{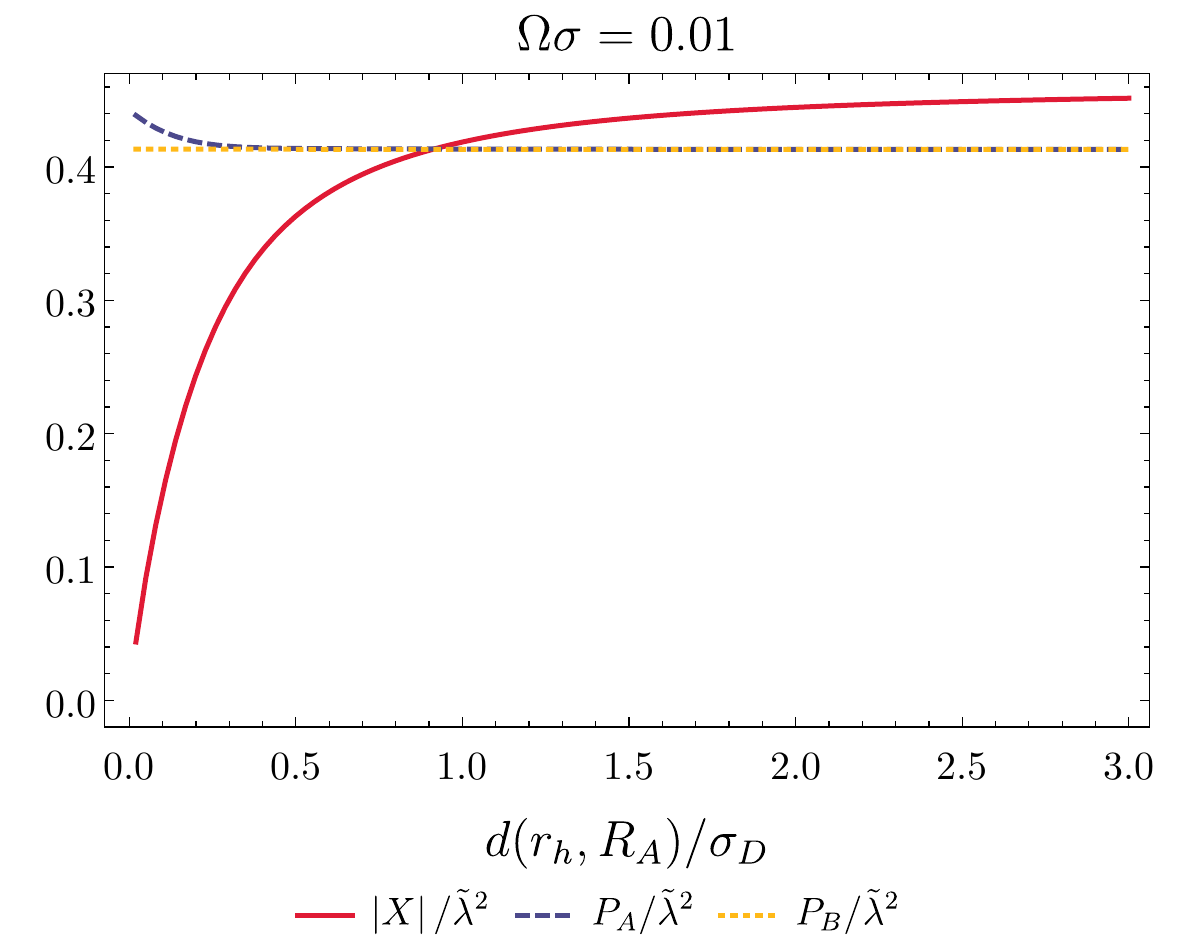}
\includegraphics[width = 0.3\textwidth]{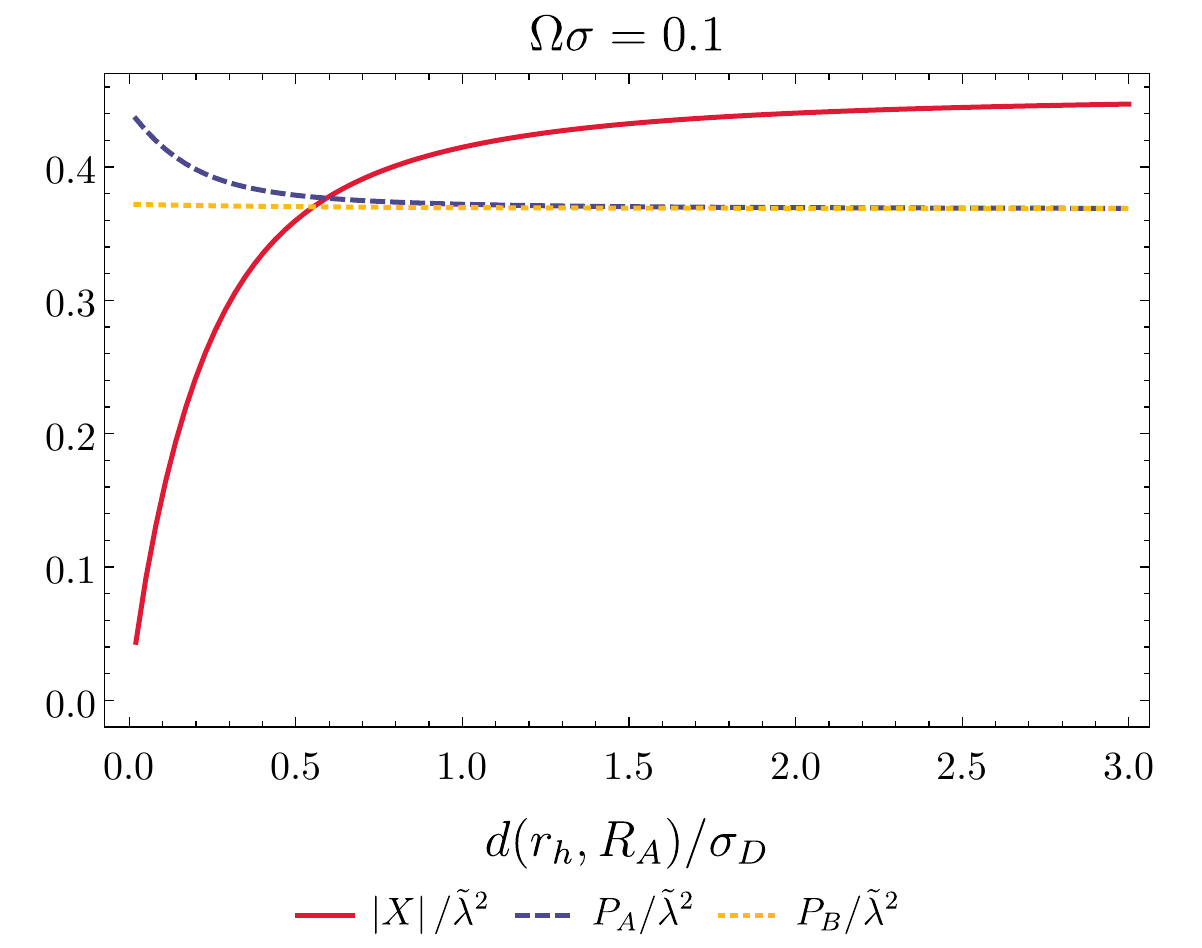}
\includegraphics[width = 0.3\textwidth]{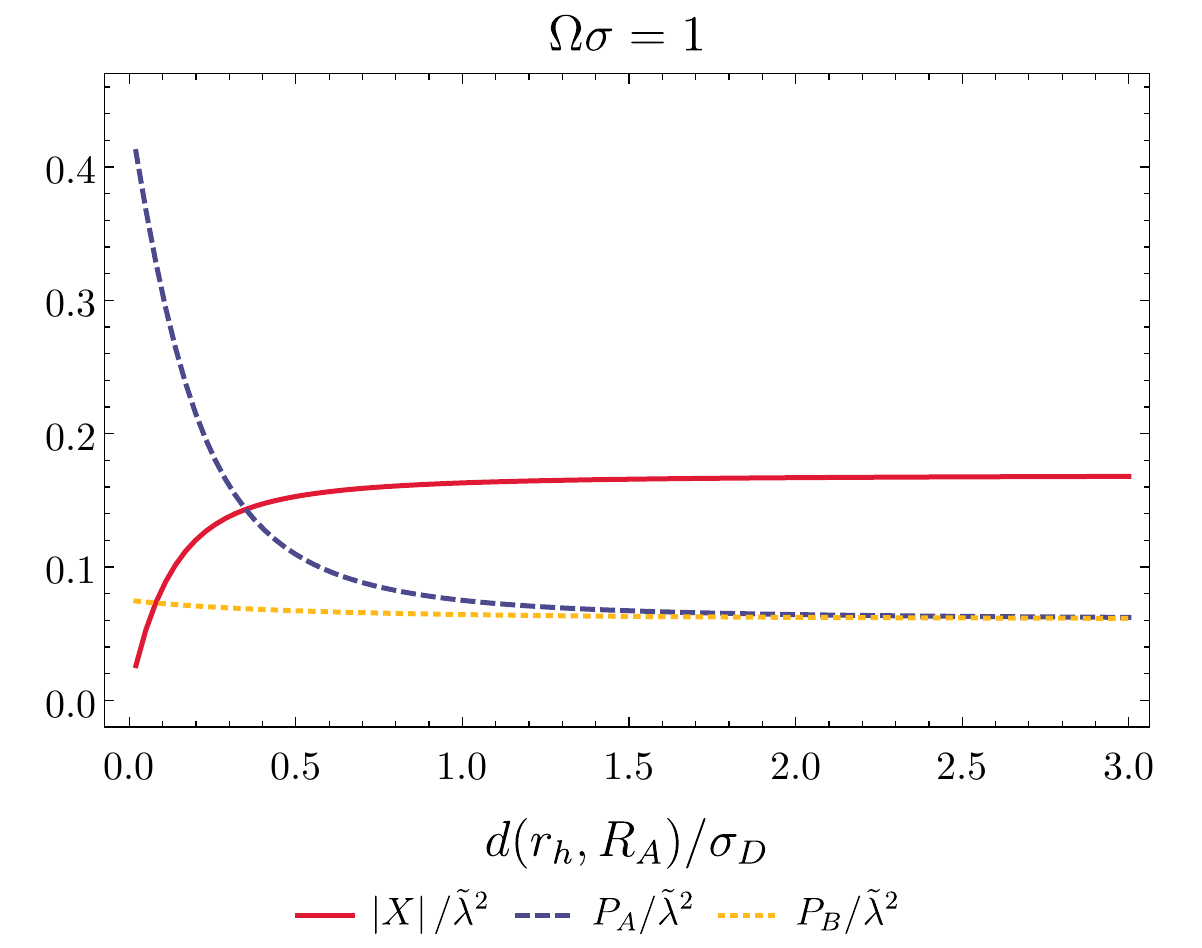}
\vspace{0.05in}
\caption{%
 The transition probabilities, $P_A$ and $P_B$, and the matrix element $|X|$ are plotted as a function of the proper distance detector $A$ is from the horizon for detectors with an energy gap of $\Omega\sigma$ of 0.01 ({\it Left}), 0.1 ({\it Center}), and 1 ({\it Right}). In all plots the detectors are separated by a proper distance of $d(R_A,R_B)/\sigma=1$, and \mbox{$\ell/\sigma = 10$ and $M=1$.}
}
\label{fig:BTZ3}
\end{figure*}

In Fig.~\ref{fig:BTZ2} we consider the two detectors, separated by a fixed proper distance, placed at various proper distances away from the horizon. For any given detector gap, a generic feature is that detectors closer to the horizon become less entangled than detectors further from the horizon. Moreover, there is a point where a ``sudden death" of entanglement harvesting occurs: detectors closer to the horizon than a certain critical proper distance, $d_{\text{death}}(r_h,R_A)$, cannot become entangled.  There are two competing phenomena that lead to this result. First, detectors closer to the horizon experience a larger local Hawking temperature~\cite{Hodfkinson:2012}, and this manifests as an increase in the transition probability of the individual detectors. Second, there is a redshift effect for the correlation term $X$, which serves to decrease $|X|$ toward zero as the detectors approach the horizon; $|X|$ can be made arbitrarily small by placing the detectors at an appropriately close distance to the horizon. This can be seen from the definition of $X$ in \eqref{defX} by changing the integration variables from the proper time of the detectors to the coordinate time, $\tau_A = \gamma_A t$ and $\tau_B = \gamma_B t'$, and noting the appearance of the redshift factors $\gamma_A$ and $\gamma_B$, which vanish as the detectors approach the horizon. Since entanglement requires that $|X| - \sqrt{P_A P_B} > 0$, each of these effects contribute to the sudden death of entanglement observed for detectors near the horizon.

The relative importance of the two effects depends on the detectors' energy gaps $\Omega$, as is further highlighted in Fig.~\ref{fig:BTZ3}. Generically, for a given proper separation between the detectors, in the limit when the detectors are far from the horizon, the values  for $|X|$ and the transition probability decrease as the detectors' energy  gaps are increased. Intuitively, this makes sense as detectors with a larger energy gap are less sensitive to fluctuations in the field. Furthermore, we see that as the detectors move toward the horizon,  the transition probability displays a larger relative change with increasing gap. The decreasing sensitivity to field fluctuations means in turn that the detectors become more sensitive to Hawking radiation.

This leads to the following interpretation. For small detector energy gaps, it is the decrease in $|X|$, due to  redshift effects, that is responsible for the sudden death of entanglement, since the transition probability remains effectively constant on the scales over which $|X|$ changes; see Fig.~\ref{fig:BTZ3}. {For larger energy gaps, the effect of the local Hawking temperature on the transition probability is more pronounced, and the sudden death is a combined effect of the redshift decreasing $|X|$ and the local Hawking temperature increasing the transition probability of the detectors.   However, in these cases the detector gap is comparable to the mass of the black hole, and therefore the gravitational backreaction cannot be safely ignored.}

In fact, a close examination of the energy gap dependence  shown in the middle plot of Fig~\ref{fig:BTZ2} reveals that there is an optimal value of the gap that pushes the point of sudden death as close to the horizon as possible. This is a consequence of the competition between the increase in the sensitivity to the Hawking effect and the overall suppression of the background values of $|X|$ and the transition probabilities. The same effect also leads  to an optimal value of $\Omega$ for  maximal   entanglement extraction from the vacuum, shown in the right plot of Fig~\ref{fig:BTZ2}.  The optimal value generically differs in each case.

We have confirmed that the results in Fig.~2 are a property of the BTZ spacetime and not simply an AdS$_3$ effect, and we have observed that the mass parameter $M$ does not affect any of the results quantitatively.  Our calculations have assumed that the detectors are turned on for the same amount of time in their own rest frame.  We have  examined switching in other frames, e.g. in the black hole frame, and find this does not  affect our basic results. Furthermore, while the  results we present are for Dirichlet boundary conditions ($\zeta = 1$), we have checked that  this choice does not qualitatively affect the behaviour we observe.

In summary, we have reported on the first investigation of the entanglement harvesting protocol in a black hole background.  We established that the amount of entanglement extracted decreases with increasing proper distance between the detectors, {commensurate with other studies \cite{Reznik:2005, Salton:2014jaa,Martin-MartinezSUS, Pozas-Kerstjens:2015, Pozas-Kerstjens:2016, Steeg:2009, Smith:2016a} probing the entanglement structure of the field.}

More surprisingly we find that black holes inhibit entanglement harvesting. As two detectors with fixed proper separation move closer to the horizon, they encounter a point of ``sudden death" where the harvested entanglement falls to zero. No entanglement can be harvested between this point and the horizon, though it is possible to minimize its proper distance from the horizon by choosing an optimal
value of the detectors' energy gap. {We find this holds even for timelike separations of the detectors, where a field-mediated interaction might be expected to increase detector entanglement.} For detectors with a small energy gap, this effect is primarily due to a decrease in the correlation term $|X|$ because of the influence horizon redshift.

This observation suggests some element of universality to our result: we suspect that this sudden death near the horizon is not a peculiarity of the BTZ spacetime, but will be present in other black holes as well, or whenever the relative red shift factor $d\tau_A/d\tau_B$ between the detectors approaches zero. We will report more extensively on the BTZ case in forthcoming work~\cite{henderForthcoming}.
\bigskip

We are grateful to Jorma Louko, Eduardo Mart{\'i}n-Mart{\'i}nez, and Keith Ng for helpful discussions. This work was supported in part by the Natural Sciences and Engineering Research Council of Canada and the Dartmouth College Society of Fellows. J. Zhang thanks the China scholarship council for financial support.

%========================================
%========================================

\bibliography{BTZpaper}

\begin{thebibliography}{36}%
\makeatletter
\providecommand \@ifxundefined [1]{%
 \@ifx{#1\undefined}
}%
\providecommand \@ifnum [1]{%
 \ifnum #1\expandafter \@firstoftwo
 \else \expandafter \@secondoftwo
 \fi
}%
\providecommand \@ifx [1]{%
 \ifx #1\expandafter \@firstoftwo
 \else \expandafter \@secondoftwo
 \fi
}%
\providecommand \natexlab [1]{#1}%
\providecommand \enquote  [1]{``#1''}%
\providecommand \bibnamefont  [1]{#1}%
\providecommand \bibfnamefont [1]{#1}%
\providecommand \citenamefont [1]{#1}%
\providecommand \href@noop [0]{\@secondoftwo}%
\providecommand \href [0]{\begingroup \@sanitize@url \@href}%
\providecommand \@href[1]{\@@startlink{#1}\@@href}%
\providecommand \@@href[1]{\endgroup#1\@@endlink}%
\providecommand \@sanitize@url [0]{\catcode `\\12\catcode `\$12\catcode
  `\&12\catcode `\#12\catcode `\^12\catcode `\_12\catcode `\%12\relax}%
\providecommand \@@startlink[1]{}%
\providecommand \@@endlink[0]{}%
\providecommand \url  [0]{\begingroup\@sanitize@url \@url }%
\providecommand \@url [1]{\endgroup\@href {#1}{\urlprefix }}%
\providecommand \urlprefix  [0]{URL }%
\providecommand \Eprint [0]{\href }%
\providecommand \doibase [0]{http://dx.doi.org/}%
\providecommand \selectlanguage [0]{\@gobble}%
\providecommand \bibinfo  [0]{\@secondoftwo}%
\providecommand \bibfield  [0]{\@secondoftwo}%
\providecommand \translation [1]{[#1]}%
\providecommand \BibitemOpen [0]{}%
\providecommand \bibitemStop [0]{}%
\providecommand \bibitemNoStop [0]{.\EOS\space}%
\providecommand \EOS [0]{\spacefactor3000\relax}%
\providecommand \BibitemShut  [1]{\csname bibitem#1\endcsname}%
\let\auto@bib@innerbib\@empty
%</preamble>
\bibitem [{\citenamefont {Osterloh}\ \emph {et~al.}(2002)\citenamefont
  {Osterloh}, \citenamefont {Amico}, \citenamefont {Falci},\ and\ \citenamefont
  {Fazio}}]{Osterloh:2002}%
  \BibitemOpen
  \bibfield  {author} {\bibinfo {author} {\bibfnamefont {A.}~\bibnamefont
  {Osterloh}}, \bibinfo {author} {\bibfnamefont {L.}~\bibnamefont {Amico}},
  \bibinfo {author} {\bibfnamefont {G.}~\bibnamefont {Falci}}, \ and\ \bibinfo
  {author} {\bibfnamefont {R.}~\bibnamefont {Fazio}},\ }\href
  {https://www.nature.com/articles/416608a} {\bibfield  {journal} {\bibinfo
  {journal} {Nature}\ }\textbf {\bibinfo {volume} {416}},\ \bibinfo {pages}
  {608} (\bibinfo {year} {2002})}\BibitemShut {NoStop}%
\bibitem [{\citenamefont {Vidal}\ \emph {et~al.}(2003)\citenamefont {Vidal},
  \citenamefont {Latorre}, \citenamefont {Rico},\ and\ \citenamefont
  {Kitaev}}]{Vidal:2003}%
  \BibitemOpen
  \bibfield  {author} {\bibinfo {author} {\bibfnamefont {G.}~\bibnamefont
  {Vidal}}, \bibinfo {author} {\bibfnamefont {J.~I.}\ \bibnamefont {Latorre}},
  \bibinfo {author} {\bibfnamefont {E.}~\bibnamefont {Rico}}, \ and\ \bibinfo
  {author} {\bibfnamefont {A.}~\bibnamefont {Kitaev}},\ }\href
  {https://journals.aps.org/prl/abstract/10.1103/PhysRevLett.90.227902}
  {\bibfield  {journal} {\bibinfo  {journal} {Phys. Rev. Lett.}\ }\textbf
  {\bibinfo {volume} {90}},\ \bibinfo {pages} {227902} (\bibinfo {year}
  {2003})}\BibitemShut {NoStop}%
\bibitem [{\citenamefont {Amico}\ \emph {et~al.}(2008)\citenamefont {Amico},
  \citenamefont {Fazio}, \citenamefont {Osterloh},\ and\ \citenamefont
  {Vedral}}]{Amico:2008}%
  \BibitemOpen
  \bibfield  {author} {\bibinfo {author} {\bibfnamefont {L.}~\bibnamefont
  {Amico}}, \bibinfo {author} {\bibfnamefont {R.}~\bibnamefont {Fazio}},
  \bibinfo {author} {\bibfnamefont {A.}~\bibnamefont {Osterloh}}, \ and\
  \bibinfo {author} {\bibfnamefont {V.}~\bibnamefont {Vedral}},\ }\href
  {https://journals.aps.org/rmp/abstract/10.1103/RevModPhys.80.517} {\bibfield
  {journal} {\bibinfo  {journal} {Rev. Mod. Phys.}\ }\textbf {\bibinfo {volume}
  {80}},\ \bibinfo {pages} {517} (\bibinfo {year} {2008})}\BibitemShut
  {NoStop}%
\bibitem [{\citenamefont {Mandel}\ and\ \citenamefont
  {Wolf}(1995)}]{Mandel:1995}%
  \BibitemOpen
  \bibfield  {author} {\bibinfo {author} {\bibfnamefont {L.}~\bibnamefont
  {Mandel}}\ and\ \bibinfo {author} {\bibfnamefont {E.}~\bibnamefont {Wolf}},\
  }\href@noop {} {\emph {\bibinfo {title} {Optical Coherence and Quantum
  Optics}}}\ (\bibinfo  {publisher} {Cambridge University Press},\ \bibinfo
  {address} {Cambridge},\ \bibinfo {year} {1995})\BibitemShut {NoStop}%
\bibitem [{\citenamefont {Weedbrook}\ \emph {et~al.}(2012)\citenamefont
  {Weedbrook}, \citenamefont {Pirandola}, \citenamefont
  {Garc\'{i}a-Patr\'{o}n}, \citenamefont {Cerf}, \citenamefont {Ralph},
  \citenamefont {Shapiro},\ and\ \citenamefont
  {Lloyd}}]{Quantum-Information:2011}%
  \BibitemOpen
  \bibfield  {author} {\bibinfo {author} {\bibfnamefont {C.}~\bibnamefont
  {Weedbrook}}, \bibinfo {author} {\bibfnamefont {S.}~\bibnamefont
  {Pirandola}}, \bibinfo {author} {\bibfnamefont {R.}~\bibnamefont
  {Garc\'{i}a-Patr\'{o}n}}, \bibinfo {author} {\bibfnamefont {N.~J.}\
  \bibnamefont {Cerf}}, \bibinfo {author} {\bibfnamefont {T.~C.}\ \bibnamefont
  {Ralph}}, \bibinfo {author} {\bibfnamefont {J.~H.}\ \bibnamefont {Shapiro}},
  \ and\ \bibinfo {author} {\bibfnamefont {S.}~\bibnamefont {Lloyd}},\ }\href
  {\doibase 10.1103/RevModPhys.84.621} {\bibfield  {journal} {\bibinfo
  {journal} {Rev. Mod. Phys.}\ }\textbf {\bibinfo {volume} {84}},\ \bibinfo
  {pages} {621} (\bibinfo {year} {2012})}\BibitemShut {NoStop}%
\bibitem [{\citenamefont {Bombelli}\ \emph {et~al.}(1986)\citenamefont
  {Bombelli}, \citenamefont {Koul}, \citenamefont {Lee},\ and\ \citenamefont
  {Sorkin}}]{Bombelli:1986}%
  \BibitemOpen
  \bibfield  {author} {\bibinfo {author} {\bibfnamefont {L.}~\bibnamefont
  {Bombelli}}, \bibinfo {author} {\bibfnamefont {R.~K.}\ \bibnamefont {Koul}},
  \bibinfo {author} {\bibfnamefont {J.}~\bibnamefont {Lee}}, \ and\ \bibinfo
  {author} {\bibfnamefont {R.~D.}\ \bibnamefont {Sorkin}},\ }\href
  {https://doi.org/10.1103/PhysRevD.34.373} {\bibfield  {journal} {\bibinfo
  {journal} {Phys. Rev. D}\ }\textbf {\bibinfo {volume} {80}},\ \bibinfo
  {pages} {373} (\bibinfo {year} {1986})}\BibitemShut {NoStop}%
\bibitem [{\citenamefont {Callan}\ and\ \citenamefont
  {Wilczek}(1994)}]{Callan:1994}%
  \BibitemOpen
  \bibfield  {author} {\bibinfo {author} {\bibfnamefont {C.}~\bibnamefont
  {Callan}}\ and\ \bibinfo {author} {\bibfnamefont {F.}~\bibnamefont
  {Wilczek}},\ }\href {\doibase 10.1016/0370-2693(94)91007-3} {\bibfield
  {journal} {\bibinfo  {journal} {Phys. Lett. B}\ }\textbf {\bibinfo {volume}
  {33}},\ \bibinfo {pages} {55} (\bibinfo {year} {1994})}\BibitemShut {NoStop}%
\bibitem [{\citenamefont {Srednicki}(1993)}]{Srednicki:1993}%
  \BibitemOpen
  \bibfield  {author} {\bibinfo {author} {\bibfnamefont {M.}~\bibnamefont
  {Srednicki}},\ }\href {\doibase 10.1103/PhysRevLett.71.666} {\bibfield
  {journal} {\bibinfo  {journal} {Phys. Rev. Lett.}\ }\textbf {\bibinfo
  {volume} {71}},\ \bibinfo {pages} {666} (\bibinfo {year} {1993})}\BibitemShut
  {NoStop}%
\bibitem [{\citenamefont {Ryu}\ and\ \citenamefont
  {Takayanagi}(2006)}]{Ryu:2006}%
  \BibitemOpen
  \bibfield  {author} {\bibinfo {author} {\bibfnamefont {S.}~\bibnamefont
  {Ryu}}\ and\ \bibinfo {author} {\bibfnamefont {T.}~\bibnamefont
  {Takayanagi}},\ }\href {\doibase 10.1103/PhysRevLett.96.181602} {\bibfield
  {journal} {\bibinfo  {journal} {Phys. Rev. Lett.}\ }\textbf {\bibinfo
  {volume} {96}},\ \bibinfo {pages} {181602} (\bibinfo {year}
  {2006})}\BibitemShut {NoStop}%
\bibitem [{\citenamefont {Summers}\ and\ \citenamefont
  {Werner}(1985)}]{Summers:1985}%
  \BibitemOpen
  \bibfield  {author} {\bibinfo {author} {\bibfnamefont {S.~J.}\ \bibnamefont
  {Summers}}\ and\ \bibinfo {author} {\bibfnamefont {R.}~\bibnamefont
  {Werner}},\ }\href
  {http://www.sciencedirect.com/science/article/pii/0375960185900933}
  {\bibfield  {journal} {\bibinfo  {journal} {Phys. Lett. A}\ }\textbf
  {\bibinfo {volume} {110}},\ \bibinfo {pages} {257} (\bibinfo {year}
  {1985})}\BibitemShut {NoStop}%
\bibitem [{\citenamefont {Summers}\ and\ \citenamefont
  {Werner}(1987{\natexlab{a}})}]{Summers:1987fn}%
  \BibitemOpen
  \bibfield  {author} {\bibinfo {author} {\bibfnamefont {S.~J.}\ \bibnamefont
  {Summers}}\ and\ \bibinfo {author} {\bibfnamefont {R.}~\bibnamefont
  {Werner}},\ }\href {\doibase 10.1063/1.527733} {\bibfield  {journal}
  {\bibinfo  {journal} {J. Math. Phys.}\ }\textbf {\bibinfo {volume} {28}},\
  \bibinfo {pages} {2440} (\bibinfo {year} {1987}{\natexlab{a}})}\BibitemShut
  {NoStop}%
%%CITATION = JMAPA,28,2440;%%
\bibitem [{\citenamefont {Summers}\ and\ \citenamefont
  {Werner}(1987{\natexlab{b}})}]{Summers:1987}%
  \BibitemOpen
  \bibfield  {author} {\bibinfo {author} {\bibfnamefont {S.~J.}\ \bibnamefont
  {Summers}}\ and\ \bibinfo {author} {\bibfnamefont {R.}~\bibnamefont
  {Werner}},\ }\href {https://doi.org/10.1063/1.527733} {\bibfield  {journal}
  {\bibinfo  {journal} {J. Math. Phys.}\ }\textbf {\bibinfo {volume} {28}},\
  \bibinfo {pages} {2448} (\bibinfo {year} {1987}{\natexlab{b}})}\BibitemShut
  {NoStop}%
\bibitem [{\citenamefont {Valentini}(1991)}]{Valentini:1991}%
  \BibitemOpen
  \bibfield  {author} {\bibinfo {author} {\bibfnamefont {A.}~\bibnamefont
  {Valentini}},\ }\href
  {http://www.sciencedirect.com/science/article/pii/0375960191909525}
  {\bibfield  {journal} {\bibinfo  {journal} {Phys. Lett. A}\ }\textbf
  {\bibinfo {volume} {153}},\ \bibinfo {pages} {321} (\bibinfo {year}
  {1991})}\BibitemShut {NoStop}%
\bibitem [{\citenamefont {Reznik}(2003)}]{Reznik:2002fz}%
  \BibitemOpen
  \bibfield  {author} {\bibinfo {author} {\bibfnamefont {B.}~\bibnamefont
  {Reznik}},\ }\href {\doibase 10.1023/A:1022875910744} {\bibfield  {journal}
  {\bibinfo  {journal} {Found. Phys.}\ }\textbf {\bibinfo {volume} {33}},\
  \bibinfo {pages} {167} (\bibinfo {year} {2003})}\BibitemShut {NoStop}%
\bibitem [{\citenamefont {Reznik}\ \emph {et~al.}(2005)\citenamefont {Reznik},
  \citenamefont {Retzker},\ and\ \citenamefont {Silman}}]{Reznik:2005}%
  \BibitemOpen
  \bibfield  {author} {\bibinfo {author} {\bibfnamefont {B.}~\bibnamefont
  {Reznik}}, \bibinfo {author} {\bibfnamefont {A.}~\bibnamefont {Retzker}}, \
  and\ \bibinfo {author} {\bibfnamefont {J.}~\bibnamefont {Silman}},\ }\href
  {http://journals.aps.org/pra/abstract/10.1103/PhysRevA.71.042104} {\bibfield
  {journal} {\bibinfo  {journal} {Phys. Rev. A}\ }\textbf {\bibinfo {volume}
  {71}},\ \bibinfo {pages} {042104} (\bibinfo {year} {2005})}\BibitemShut
  {NoStop}%
\bibitem [{\citenamefont {Salton}\ \emph {et~al.}(2015)\citenamefont {Salton},
  \citenamefont {Mann},\ and\ \citenamefont {Menicucci}}]{Salton:2014jaa}%
  \BibitemOpen
  \bibfield  {author} {\bibinfo {author} {\bibfnamefont {G.}~\bibnamefont
  {Salton}}, \bibinfo {author} {\bibfnamefont {R.~B.}\ \bibnamefont {Mann}}, \
  and\ \bibinfo {author} {\bibfnamefont {N.~C.}\ \bibnamefont {Menicucci}},\
  }\href {\doibase 10.1088/1367-2630/17/3/035001} {\bibfield  {journal}
  {\bibinfo  {journal} {New J. Phys.}\ }\textbf {\bibinfo {volume} {17}},\
  \bibinfo {pages} {035001} (\bibinfo {year} {2015})}\BibitemShut {NoStop}%
\bibitem [{\citenamefont {Mart\'{i}n-Mart\'{i}nez}\ \emph
  {et~al.}(2013{\natexlab{a}})\citenamefont {Mart\'{i}n-Mart\'{i}nez},
  \citenamefont {Brown}, \citenamefont {Donnelly},\ and\ \citenamefont
  {Kempf}}]{Martin-MartinezSUS}%
  \BibitemOpen
  \bibfield  {author} {\bibinfo {author} {\bibfnamefont {E.}~\bibnamefont
  {Mart\'{i}n-Mart\'{i}nez}}, \bibinfo {author} {\bibfnamefont {E.~G.}\
  \bibnamefont {Brown}}, \bibinfo {author} {\bibfnamefont {W.}~\bibnamefont
  {Donnelly}}, \ and\ \bibinfo {author} {\bibfnamefont {A.}~\bibnamefont
  {Kempf}},\ }\href {\doibase 10.1103/PhysRevA.88.052310} {\bibfield  {journal}
  {\bibinfo  {journal} {Phys. Rev. A}\ }\textbf {\bibinfo {volume} {88}},\
  \bibinfo {pages} {052310} (\bibinfo {year} {2013}{\natexlab{a}})}\BibitemShut
  {NoStop}%
\bibitem [{\citenamefont {Pozas-Kerstjens}\ and\ \citenamefont
  {Mart{\'\i}n-Mart{\'\i}nez}(2015)}]{Pozas-Kerstjens:2015}%
  \BibitemOpen
  \bibfield  {author} {\bibinfo {author} {\bibfnamefont {A.}~\bibnamefont
  {Pozas-Kerstjens}}\ and\ \bibinfo {author} {\bibfnamefont {E.}~\bibnamefont
  {Mart{\'\i}n-Mart{\'\i}nez}},\ }\href {\doibase 10.1103/PhysRevD.92.064042}
  {\bibfield  {journal} {\bibinfo  {journal} {Phys. Rev. D}\ }\textbf {\bibinfo
  {volume} {92}},\ \bibinfo {pages} {064042} (\bibinfo {year}
  {2015})}\BibitemShut {NoStop}%
\bibitem [{\citenamefont {Pozas-Kerstjens}\ and\ \citenamefont
  {Mart\'{i}n-Mart\'{i}nez}(2016)}]{Pozas-Kerstjens:2016}%
  \BibitemOpen
  \bibfield  {author} {\bibinfo {author} {\bibfnamefont {A.}~\bibnamefont
  {Pozas-Kerstjens}}\ and\ \bibinfo {author} {\bibfnamefont {E.}~\bibnamefont
  {Mart\'{i}n-Mart\'{i}nez}},\ }\href {\doibase 10.1103/PhysRevD.94.064074}
  {\bibfield  {journal} {\bibinfo  {journal} {Phys. Rev. D}\ }\textbf {\bibinfo
  {volume} {94}},\ \bibinfo {pages} {064074} (\bibinfo {year}
  {2016})}\BibitemShut {NoStop}%
\bibitem [{\citenamefont {{Ver Steeg}}\ and\ \citenamefont
  {Menicucci}(2009)}]{Steeg:2009}%
  \BibitemOpen
  \bibfield  {author} {\bibinfo {author} {\bibfnamefont {G.}~\bibnamefont {{Ver
  Steeg}}}\ and\ \bibinfo {author} {\bibfnamefont {N.~C.}\ \bibnamefont
  {Menicucci}},\ }\href
  {http://journals.aps.org/prd/abstract/10.1103/PhysRevD.79.044027} {\bibfield
  {journal} {\bibinfo  {journal} {Phys. Rev. D}\ }\textbf {\bibinfo {volume}
  {79}},\ \bibinfo {pages} {044027} (\bibinfo {year} {2009})}\BibitemShut
  {NoStop}%
\bibitem [{\citenamefont {Mart\'{i}n-Mart\'{i}nez}\ \emph
  {et~al.}(2016)\citenamefont {Mart\'{i}n-Mart\'{i}nez}, \citenamefont {R. {\
  H. Smith}},\ and\ \citenamefont {Terno}}]{Smith:2016a}%
  \BibitemOpen
  \bibfield  {author} {\bibinfo {author} {\bibfnamefont {E.}~\bibnamefont
  {Mart\'{i}n-Mart\'{i}nez}}, \bibinfo {author} {\bibfnamefont
  {A.}~\bibnamefont {R. {\ H. Smith}}}, \ and\ \bibinfo {author}
  {\bibfnamefont {D.~R.}\ \bibnamefont {Terno}},\ }\href
  {https://journals.aps.org/prd/abstract/10.1103/PhysRevD.93.044001} {\bibfield
   {journal} {\bibinfo  {journal} {Phys. Rev. D}\ }\textbf {\bibinfo {volume}
  {93}},\ \bibinfo {pages} {044001} (\bibinfo {year} {2016})}\BibitemShut
  {NoStop}%
\bibitem [{\citenamefont {Unruh}(1976)}]{Unruh:1976}%
  \BibitemOpen
  \bibfield  {author} {\bibinfo {author} {\bibfnamefont {W.~G.}\ \bibnamefont
  {Unruh}},\ }\href
  {http://journals.aps.org/prd/abstract/10.1103/PhysRevD.14.870} {\bibfield
  {journal} {\bibinfo  {journal} {Phys. Rev. D}\ }\textbf {\bibinfo {volume}
  {14}},\ \bibinfo {pages} {870} (\bibinfo {year} {1976})}\BibitemShut
  {NoStop}%
\bibitem [{\citenamefont {DeWitt}(1979)}]{DeWitt:1979}%
  \BibitemOpen
  \bibfield  {author} {\bibinfo {author} {\bibfnamefont {B.~S.}\ \bibnamefont
  {DeWitt}},\ }\enquote {\bibinfo {title} {{General Relativity: An Einsten
  Centenary Survey}},}\ \ (\bibinfo  {publisher} {Cambridge University Press},\
  \bibinfo {address} {Cambridge},\ \bibinfo {year} {1979})\ Chap.\ \bibinfo
  {chapter} {{Quantum Gravity: The New Synthesis}}, pp.\ \bibinfo {pages}
  {680--745}\BibitemShut {NoStop}%
\bibitem [{\citenamefont {Mart\'{i}n-Mart\'{i}nez}\ \emph
  {et~al.}(2013{\natexlab{b}})\citenamefont {Mart\'{i}n-Mart\'{i}nez},
  \citenamefont {Montero},\ and\ \citenamefont {del
  Rey}}]{Martin-Martinez2013}%
  \BibitemOpen
  \bibfield  {author} {\bibinfo {author} {\bibfnamefont {E.}~\bibnamefont
  {Mart\'{i}n-Mart\'{i}nez}}, \bibinfo {author} {\bibfnamefont
  {M.}~\bibnamefont {Montero}}, \ and\ \bibinfo {author} {\bibfnamefont
  {M.}~\bibnamefont {del Rey}},\ }\href {\doibase 10.1103/PhysRevD.87.064038}
  {\bibfield  {journal} {\bibinfo  {journal} {Phys. Rev. D}\ }\textbf {\bibinfo
  {volume} {87}},\ \bibinfo {pages} {064038} (\bibinfo {year}
  {2013}{\natexlab{b}})}\BibitemShut {NoStop}%
\bibitem [{\citenamefont {Alhambra}\ \emph {et~al.}(2014)\citenamefont
  {Alhambra}, \citenamefont {Kempf},\ and\ \citenamefont
  {Mart\'in-Mart\'inez}}]{Alvaro}%
  \BibitemOpen
  \bibfield  {author} {\bibinfo {author} {\bibfnamefont {A.~M.}\ \bibnamefont
  {Alhambra}}, \bibinfo {author} {\bibfnamefont {A.}~\bibnamefont {Kempf}}, \
  and\ \bibinfo {author} {\bibfnamefont {E.}~\bibnamefont
  {Mart\'in-Mart\'inez}},\ }\href {\doibase 10.1103/PhysRevA.89.033835}
  {\bibfield  {journal} {\bibinfo  {journal} {Phys. Rev. A}\ }\textbf {\bibinfo
  {volume} {89}},\ \bibinfo {pages} {033835} (\bibinfo {year}
  {2014})}\BibitemShut {NoStop}%
\bibitem [{\citenamefont {Smith}(2017)}]{Smith:2017b}%
  \BibitemOpen
  \bibfield  {author} {\bibinfo {author} {\bibfnamefont {A.~R.~H.}\
  \bibnamefont {Smith}},\ }\emph {\bibinfo {title} {Detectors, Reference
  Frames, and Time}},\ \href {https://uwspace.uwaterloo.ca/handle/10012/12618}
  {Ph.D. thesis},\ \bibinfo  {school} {University of Waterloo and Macquarie
  University} (\bibinfo {year} {2017})\BibitemShut {NoStop}%
\bibitem [{Note1()}]{Note1}%
  \BibitemOpen
  \bibinfo {note} {Other literature on entanglement harvesting has used the
  negativity as a measure of entanglement, which when evaluated for the state
  given in \protect \textup {\hbox {\mathsurround \z@ \protect \normalfont
  (\ignorespaces \ref {FinsalState2}\unskip \@@italiccorr )}} to leading order
  yields {\relax \protect \fontsize {7}{8}\protect \selectfont \begin {align*}
  \protect \mathcal {N}(\rho _{AB}) &= \protect \qopname \relax m{max}\protect
  \tmspace -\thinmuskip {.1667em}\left [ \protect \tmspace +\thinmuskip
  {.1667em} 0, \protect \tmspace +\thinmuskip {.1667em} \protect \sqrt { \left
  | X \right |^2 + \left (\protect \frac {P_A-P_B}{2}\right )^2} -\protect
  \frac {P_A+P_B}{2}\protect \tmspace +\thinmuskip {.1667em} \right ] . \end
  {align*} } However, unlike concurrence, the negativity is not a simple
  difference between a nonlocal and a local term. It is for this reason we
  employ the concurrence in this article.}\BibitemShut {Stop}%
\bibitem [{\citenamefont {Wootters}(1998)}]{Wootters:1997id}%
  \BibitemOpen
  \bibfield  {author} {\bibinfo {author} {\bibfnamefont {W.~K.}\ \bibnamefont
  {Wootters}},\ }\href {\doibase 10.1103/PhysRevLett.80.2245} {\bibfield
  {journal} {\bibinfo  {journal} {Phys. Rev. Lett.}\ }\textbf {\bibinfo
  {volume} {80}},\ \bibinfo {pages} {2245} (\bibinfo {year}
  {1998})}\BibitemShut {NoStop}%
\bibitem [{\citenamefont {{Ba\~{n}ados}}\ \emph {et~al.}(1992)\citenamefont
  {{Ba\~{n}ados}}, \citenamefont {Teitelboim},\ and\ \citenamefont
  {Zanelli}}]{Banados:1992}%
  \BibitemOpen
  \bibfield  {author} {\bibinfo {author} {\bibfnamefont {M.}~\bibnamefont
  {{Ba\~{n}ados}}}, \bibinfo {author} {\bibfnamefont {C.}~\bibnamefont
  {Teitelboim}}, \ and\ \bibinfo {author} {\bibfnamefont {J.}~\bibnamefont
  {Zanelli}},\ }\href {\doibase 10.1103/PhysRevLett.69.1849} {\bibfield
  {journal} {\bibinfo  {journal} {Phys. Rev. Lett.}\ }\textbf {\bibinfo
  {volume} {69}},\ \bibinfo {pages} {1849} (\bibinfo {year}
  {1992})}\BibitemShut {NoStop}%
\bibitem [{\citenamefont {{Ba\~{n}ados}}\ \emph {et~al.}(1993)\citenamefont
  {{Ba\~{n}ados}}, \citenamefont {Henneaux}, \citenamefont {Teitelboim},\ and\
  \citenamefont {Zanelli}}]{Banados:1993}%
  \BibitemOpen
  \bibfield  {author} {\bibinfo {author} {\bibfnamefont {M.}~\bibnamefont
  {{Ba\~{n}ados}}}, \bibinfo {author} {\bibfnamefont {M.}~\bibnamefont
  {Henneaux}}, \bibinfo {author} {\bibfnamefont {C.}~\bibnamefont
  {Teitelboim}}, \ and\ \bibinfo {author} {\bibfnamefont {J.}~\bibnamefont
  {Zanelli}},\ }\href {\doibase 10.1103/PhysRevD.48.1506} {\bibfield  {journal}
  {\bibinfo  {journal} {Phys. Rev. D}\ }\textbf {\bibinfo {volume} {48}},\
  \bibinfo {pages} {1506} (\bibinfo {year} {1993})}\BibitemShut {NoStop}%
\bibitem [{\citenamefont {Hodgkinson}\ and\ \citenamefont
  {Louko}(2012)}]{Hodfkinson:2012}%
  \BibitemOpen
  \bibfield  {author} {\bibinfo {author} {\bibfnamefont {L.}~\bibnamefont
  {Hodgkinson}}\ and\ \bibinfo {author} {\bibfnamefont {J.}~\bibnamefont
  {Louko}},\ }\href {\doibase 10.1103/PhysRevD.86.064031} {\bibfield  {journal}
  {\bibinfo  {journal} {Phys. Rev. D}\ }\textbf {\bibinfo {volume} {86}},\
  \bibinfo {pages} {064031} (\bibinfo {year} {2012})}\BibitemShut {NoStop}%
\bibitem [{\citenamefont {Hodgkinson}\ and\ \citenamefont
  {Louko}()}]{Hodfkinson:2012a}%
  \BibitemOpen
  \bibfield  {author} {\bibinfo {author} {\bibfnamefont {L.}~\bibnamefont
  {Hodgkinson}}\ and\ \bibinfo {author} {\bibfnamefont {J.}~\bibnamefont
  {Louko}},\ }\href {http://arxiv.org/abs/1208.3165} {}\bibinfo {note}
  {{a}rXiv:gr-qc/1208.3165}\BibitemShut {NoStop}%
\bibitem [{\citenamefont {Smith}\ and\ \citenamefont
  {Mann}(2014)}]{Smith:2014}%
  \BibitemOpen
  \bibfield  {author} {\bibinfo {author} {\bibfnamefont {A.~R.~H.}\
  \bibnamefont {Smith}}\ and\ \bibinfo {author} {\bibfnamefont {R.~B.}\
  \bibnamefont {Mann}},\ }\href {\doibase 10.1088/0264-9381/31/8/082001}
  {\bibfield  {journal} {\bibinfo  {journal} {Class. Quant. Grav.}\ }\textbf
  {\bibinfo {volume} {31}},\ \bibinfo {pages} {082001} (\bibinfo {year}
  {2014})}\BibitemShut {NoStop}%
\bibitem [{\citenamefont {Lifschytz}\ and\ \citenamefont
  {Ortiz}(1994)}]{Lifschytz:1994}%
  \BibitemOpen
  \bibfield  {author} {\bibinfo {author} {\bibfnamefont {G.}~\bibnamefont
  {Lifschytz}}\ and\ \bibinfo {author} {\bibfnamefont {M.}~\bibnamefont
  {Ortiz}},\ }\href {\doibase 10.1103/PhysRevD.49.1929} {\bibfield  {journal}
  {\bibinfo  {journal} {Phys. Rev. D}\ }\textbf {\bibinfo {volume} {49}},\
  \bibinfo {pages} {1929} (\bibinfo {year} {1994})}\BibitemShut {NoStop}%
\bibitem [{\citenamefont {Carlip}(2003)}]{Carlip:2003}%
  \BibitemOpen
  \bibfield  {author} {\bibinfo {author} {\bibfnamefont {S.}~\bibnamefont
  {Carlip}},\ }\href@noop {} {\emph {\bibinfo {title} {Quantum Gravity in 2+1
  Dimensions}}}\ (\bibinfo  {publisher} {Cambridge University Press},\ \bibinfo
  {address} {Cambridge},\ \bibinfo {year} {2003})\BibitemShut {NoStop}%
\bibitem [{\citenamefont {Henderson}\ \emph {et~al.}(2018)\citenamefont
  {Henderson}, \citenamefont {Hennigar}, \citenamefont {Mann}, \citenamefont
  {Smith},\ and\ \citenamefont {Zhang}}]{henderForthcoming}%
  \BibitemOpen
  \bibfield  {author} {\bibinfo {author} {\bibfnamefont {L.~J.}\ \bibnamefont
  {Henderson}}, \bibinfo {author} {\bibfnamefont {R.~A.}\ \bibnamefont
  {Hennigar}}, \bibinfo {author} {\bibfnamefont {R.~B.}\ \bibnamefont {Mann}},
  \bibinfo {author} {\bibfnamefont {A.~R.~H.}\ \bibnamefont {Smith}}, \ and\
  \bibinfo {author} {\bibfnamefont {J.}~\bibnamefont {Zhang}},\ }\href@noop {}
  {} (\bibinfo {year} {2018}),\ \bibinfo {note} {in preparation}\BibitemShut
  {NoStop}%
\end{thebibliography}%

\onecolumngrid
\appendix

\section{Supplemental Material}

The explicit form the transition probabilities used in the numerical evaluation of the concurrence is

\begin{align*}
  P_D &= \kappa_D\int_{-\infty}^{\infty} dy\ \frac{\exp\big[-\sigma^2\big(y-\Omega\big)^2\big]}{\exp\big(y/T_D\big)+1} -\zeta K_D\int_{0}^{\infty} dy\  {\rm{Re}}\Bigg[\frac{\exp\big(-a_D y^2\big)\exp\big(-i \beta_D y\big)}{\sqrt{\cosh\big(\alpha_{D,0}^+\big)-\cosh(y)}}\Bigg]\\
  &\phantom{=} + 2\sum_{n=1}^{\infty} \Bigg\{K_D\int_{0}^{\infty} dy\ {\rm{Re}}\Bigg[ \frac{\exp\big(-a_D y^2\big)\exp\big(-i \beta_D y\big)}{\sqrt{\cosh\big(\alpha_{D,n}^-\big)-\cosh(y)}}\Bigg] - \zeta K_D\int_{0}^{\infty} dy\ {\rm{Re}}\Bigg[\frac{\exp\big(-a_D y^2\big)\exp\big(-i \beta_D y\big)}{\sqrt{\cosh\big(\alpha_{D,n}^+\big)-\cosh(y)}}\Bigg]\Bigg\},
\end{align*}
where
\begin{align*}
  \kappa_D &\ce \frac{\lambda^2\sigma^2}{2}, &
  K_D &\ce \frac{\lambda^2\sigma}{2\sqrt{2\pi}}, \\
  T_D &\ce \frac{r_h}{2\pi \ell\sqrt{R_D^2-r_h^2}}, &
  a_D &\ce \frac{\gamma_D^2 \ell^4}{4\sigma^2r_h^2},\\
  \beta_D &\ce \frac{\gamma_D\Omega\ell^2}{r_h}, &
  \alpha_{D,n}^{\mp} &\ce \arccosh\left[\frac{r_h^2}{R_D^2-r_h^2}\left(\frac{R_D^2}{r_h^2}\cosh\left(\frac{r_h}{\ell}2\pi n \right)\mp1\right)\right].
\end{align*}

The explicit form the matrix element $X$ used in the numerical evaluation of the concurrence is

\begin{align*}
  X &= -\sum_{n=-\infty}^{\infty} \Bigg[K_X\int_{0}^{\infty} dy\ \frac{\exp\big(-a_X y^2\big)\cos\big(\beta_X y\big)}{\sqrt{\cosh\big(\alpha_{X,n}^{-}\big)-\cosh(y)}}-\zeta K_X\int_{0}^{\infty} dy\ \frac{\exp\big(-a_X y^2\big)\cos\big(\beta_X y\big)}{\sqrt{\cosh\big(\alpha_{X,n}^{+}\big)-\cosh(y)}}\Bigg],
\end{align*}
where
\begin{align*}
  K_X &\colonequals \frac{\lambda^2\sigma}{2\sqrt{\pi}}\sqrt{\frac{\gamma_A\gamma_B}{\gamma_A^2+\gamma_B^2}}\exp\left(-\frac{\sigma^2\Omega^2}{2}\frac{\left(\gamma_A+\gamma_B\right)^2}{\gamma_A^2+\gamma_B^2}\right), &
  a_X &\colonequals \frac{1}{2\sigma^2}\frac{\gamma_A^2\gamma_B^2}{\gamma_A^2+\gamma_B^2}\frac{\ell^4}{r_h^2}, \\
  \beta_X &\colonequals \Omega\frac{\gamma_A\gamma_B\left(\gamma_A-\gamma_B\right)}{\gamma_A^2+\gamma_B^2}\frac{\ell^2}{r_h}, &
  \alpha_{X,n}^{\mp} &:= \arccosh\left[\frac{r_h^2}{\ell^2\gamma_A\gamma_B}\left(\frac{R_AR_B}{r_h^2}\cosh\left(\frac{r_h}{\ell}\left(\Delta\phi+2\pi n\right)\right)\mp1\right)\right].
\end{align*}

\end{document}